\documentclass{PoS}

\PoS{PoS(LAT2005)096}

\title{Heavy-light mesons with domain wall fermions }

\ShortTitle{Heavy-light mesons with DWF}

\author{\speaker{Shigemi Ohta}
\\
Inst.\ Particle and Nuclear Studies, KEK and Physics Department, Sokendai,
Tsukuba, Ibaraki 305-0801, Japan and RIKEN BNL Research Center, BNL, Upton, NY 11973, USA\\
E-mail: \email{shigemi.ohta@kek.jp}}

\author{HueyWen Lin\\
Physics Department, Columbia University, New York, NY 10027, USA\\
E-mail: \email{hwlin@physics.columbia.edu}}

\author{Norikazu Yamada\\
Inst.\ Particle and Nuclear Studies, KEK and Physics Department, Sokendai,
Tsukuba, Ibaraki 305-0801, Japan\\
E-mail: \email{norikazu.yamada@kek.jp}}

\author{[RBC Collaboration]}

\abstract{We present RBC heavy-light meson spectroscopy with quenched DBW2 gauge configurations at lattice cutoff \(a^{-1}\) of about 3 GeV.  Both heavy and light quarks are described by domain-wall fermions (DWF).  The heavy quark mass ranges between 0.1 and 0.4 lattice units, covering charm.  The light quark mass ranges between 0.008 and 0.04, covering strange.  In particular, we discuss charmed (\(D\) and \(D^{*}\)) and charm-strange (\(D_s\) and \(D_{sJ}\)) mesons with spin-parity \(J^P= 0^\pm\) and \(1^\pm\).  The preliminary results indicate that DWF describe charm on the quenched DBW2 ensemble at this cutoff.  The masses of the \(J^{P}=0^{\pm}\) and \(1^{\pm}\) \(D\), \(D^{*}\), \(D_{s}\) and \(D_{sJ}\) meson states are well reproduced to within a few \%; their parity splitting, \(\Delta_{J}\), are better reproduced than previous works, with only 10-20 \% over estimations; the experimental observation that the splitting for non-strange states, \(\Delta_{ud}\), is bigger than that for strange states, \(\Delta_{s}\), is reproduced as well; but the hyperfine splittings are only 60-65 \% reproduced.  Regarding the depenence on heavy quark mass, \(\Delta_{J=0}\) and \(\Delta_{J=1}\) are degenerate for \(m_{\rm heavy} a >\) 0.2-0.3; \(\Delta_{J=0}\) increases as \(m_{\rm heavy}\) decreases further while \(\Delta_{J=1}\) does not.}

\FullConference{XXIIIrd International Symposium on Lattice Field Theory\\
		 25-30 July 2005\\
		 Trinity College, Dublin, Ireland}

\begin{document}

\section{Introduction}

We report preliminary results from a RIKEN-BNL-Columbia (RBC) lattice numerical calculation on charmed meson spectroscopy with domain-wall fermions (DWF) \cite{Kaplan:1992bt} quarks and a quenched gauge ensemble with the rectangular improved DBW2 (Doubly-Blocked Wilson 2) \cite{Takaishi:1996xj} action at the lattice cutoff \(a^{-1}\) of about 3 GeV.  All the quarks including charm are described by DWF.

There are two main motivations: 1) By implementing charm as DWF, difficulties arising from lacking the Glashow-Illiopoulos-Miani mechanism can be avoided, there by making ``charm-in'' lattice calculations of hadron electroweak transition easier.  This calculation serves as a test for such projects.  2) Recent discoveries of \(D_{sJ}\) mesons by the B factory experiments made comparison of lattice calculations and experiments more interesting.  Planned experiments at \(\tau\)-charm factories will make this more accurate and challenging.

In Table \ref{tab:experiment} we summarize low-lying chamed (\(D\) and \(D^{*}\)) and charm-strange (\(D_{s}\) and \(D_{sJ}\)) mesons mass: spin-pariity \(J^{P}=0^{\mp}\) and \(1^{\mp}\) states are known.
\begin{table}[b]
\begin{center}
\hfill
\begin{tabular}{ll}
\multicolumn{1}{c}{\(J^{P}\)}&\multicolumn{1}{c}{mass (MeV)}\\
\hline\hline
\(D^{\pm}(0^{-})\)&1869.4(5)\\
\(D^{*\pm}(1^{-})\)&2010.0(5)\\
\(D^{*}_{0}(0^{+}?)\)&2308(17)(15)(28)\\
\(D'_{1}(1^{+})\)&2427(26)(20)(15)\\
\end{tabular}
\hfill
\begin{tabular}{ll}

\multicolumn{1}{c}{\(J^{P}\)}&\multicolumn{1}{c}{mass (MeV)}\\
\hline\hline
\(D^{\pm}_{s}(0^{-})\)&1968.3(5)\\
\(D^{*\pm}_{s}(1^{-}?)\)&2112.1(7)\\
\(D^{*\pm}_{s0}(0^{+})\)&2317.4(9)\\
\(D^{*\pm}_{s1}(1^{+})\)&2459.3(1.3)\\
\end{tabular}
\hfill
\end{center}
\caption{Experimental values of low-lying \(D\) and \(D^{*}\) (left) and \(D_{s}\) and \(D_{sJ}\) (right) mass.}
\label{tab:experiment}
\end{table}
From these we identify the following objectives for the present calculation to explain:
The splitting between parity partners, \(\Delta_{J}=m_{D_{l}}(J^{+})-m_{D_{l}}(J^{-})\) (see Table \ref{tab:paritysplitting}.)
\begin{table}[t]
\begin{center}
\begin{tabular}{cll}
\multicolumn{1}{c}{light quark}&\multicolumn{1}{c}{\(0^{+}-0^{-}\) (MeV)}&\multicolumn{1}{c}{\(1^{+}-1^{-}\) (MeV)}\\
\hline\hline
\(s\) & 349.1 & 347.2\\
\(ud\) & 439 & 417 \\
\end{tabular}
\end{center}
\caption{Mass splitting, \(\Delta_{lJ}\), between parity partners.}
\label{tab:paritysplitting}
\end{table}
A) They appear insensitive to the spin \(J\),
\[
\Delta_{l, J=0} \sim \Delta_{l, J=1}.
\]
Here the subscript \(l\) refers to the light quark flavor, either non-strange (\(ud\)) or strange (\(s\)).
B) On the other hand they seem dependent on \(m_{l}\),
\[
\Delta_{ud} > \Delta_{s}.
\]
C) The hyperfine splitting \(\Delta_{\rm hf}=m_{D_{l}}(1^{-})-m_{D_{l}}(0^{-})\) or \(m_{D_{l}}(1^{+})-m_{D_{l}}(0^{+})\) (see Table \ref{tab:hyperfinesplitting}.)
\begin{table}[b]
\begin{center}
\begin{tabular}{cll}
\multicolumn{1}{c}{light quark}&\multicolumn{1}{c}{\(1^{-}-0^{-}\) (MeV)}&\multicolumn{1}{c}{\(1^{+}-0^{+}\) (MeV)}\\
\hline\hline
\(s\) & 143.8 & 141.9\\
\(ud\) & 140.6 & 119 \\
\end{tabular}
\end{center}
\caption{Hyperfine splitting \(\Delta_{\rm hf}\)}
\label{tab:hyperfinesplitting}
\end{table}
They seem independent of \(m_{l}\) or parity.

Before the discoveries of \(D_{sJ}\) mesons, Bardeen, Eichten and Hill \cite{Bardeen:2003kt}, based on their \(SU(3)_{L}\times SU(3)_{R}\) model for heavy-light meson systems, made a set of predictions regarding the \(D\) and \(D_{s}\) spectroscopy: that the parity splitting \(\Delta_{lJ}\) would be independent of the spin \(J\) and that it is weakly dependent on the heavy quark mass.  They also made an interesting prediction that a Goleberger-Treiman like relation should hold in the pion emission from excited \(D^{*}\) or \(D_{sJ}\) states: \(g_{\pi} = \Delta_{lJ}/f_{\pi}\), though this is not in the scope of current study.  Nowak, Rho and Zahed \cite{Nowak:2003ra} made similar predictions based on a different model.  More recently Becirevic, Fajfer and Prelovsek \cite{Becirevic:2004uv}, based on yet another model, discussed it is difficult to understand the experimental observation that \(\Delta_{ud} > \Delta_{s}\).

By now many lattice numerical studies \cite{Boyle:1997rk, Hein:2000qu, Bali:2003jv, diPierro:2003iw, Green:2003zz, Dougall:2003hv} have been conducted on these and related meson states.  To summarize their results: a) The majority \cite{Boyle:1997rk, Hein:2000qu, Bali:2003jv, diPierro:2003iw} have overestimated the parity splittings, \(\Delta_{J}\), with values typically about 500 MeV.  b) A quenched calculation underestimated the hyperfine splittings reproducing only about half of the experimental values \cite{Boyle:1997rk}.  This is considered a common pathology for quenched calculations.  c) Some success was observed in reproducing the degeneracy of \(\Delta_{J=0}\sim \Delta_{J=1}\) \cite{Dougall:2003hv}.
The majority of these lattice studies employ static or non-relativistic description of the heavy charm quark, with the exception of ref.\ \cite{diPierro:2003iw} which employs the Fermilab method. In contrast, in this study we explore the possibility of propagating the charm quark as a domain-wall fermion on a high-cutoff lattice.  Needless to say, we exploit the good chiral symmetry of DWF for the light quarks.

\section{Numerical Method}

The quenched DBW2 gauge ensemble we use in this study is described in a previous RBC publication \cite{Aoki:2005ga}.  103 configurations of the total 106 described there are used here.  The gauge coupling is set at \(\beta=1.22\).  The lattice volume is \(24^{3}\times 48\) and corresponds to about \(({\rm 1.6 fm})^{3}\) spatial box as the lattice cutoff measured by the \(\rho\)-meson mass is 2.914(54) GeV.  We will use this cutoff estimate throughout this report.  If we use the static quark potential instead, we obtain 3.07 GeV \cite{Koichi}.  The difference gives an estimate of quenched systematics.

All the four quark flavors are described as domain-wall fermions (see, for example, ref.\ \cite{Blum:2000kn} for the details of numerical implementation.)  We set the fifth-dimension lattice extent as \(L_{s}=10\), the domain wall height \(M_{5}=1.65\).  The propagators are calculated with ``periodic+antiperiodic'' boundary condition in time, so the hadron propagators are periodic with a period of 96.

For the light quark mass, five bare values of 0.008, 0.016, 0.024, 0.032 and 0.040 are used.  From the calculated light-light pseudoscalar mass, the strange quark mass is known to be 0.0295(14) in lattice units \cite{Aoki:2005ga}.

\begin{figure}[t]
\begin{center}
\includegraphics[width=.45\textwidth]{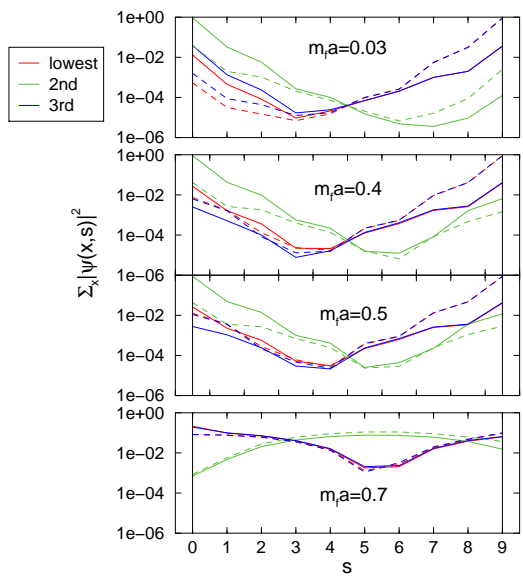}
\end{center}
\caption{Four-dimensional norms of the low-lying DWF eigenmodes.  Courtesy: Jun-Ichi Noaki, RBC Collaboration.}
\label{fig:5Dnorm}
\end{figure}

For the heavy quark mass, five values of 0.1, 0.2, 0.3, 0.4 and 0.5 are used.  Here a possible problem is that the domain wall fermion with explicit mass of \(O(1)\) may no longer be localized to the domain wall.  Jun Noaki of the RBC Collaboration investigated this issue by looking at the dependence on the fifth coordinate, \(s\), of the four-dimensional norms of the low-lying DWF eigenmodes.  As is presented in Figure \ref{fig:5Dnorm},
the DWF are localized to the wall for explicit mass values smaller than about 0.5.

\section{Physics Results}

\begin{figure}[b]
\begin{center}
\includegraphics[width=.5\textwidth]{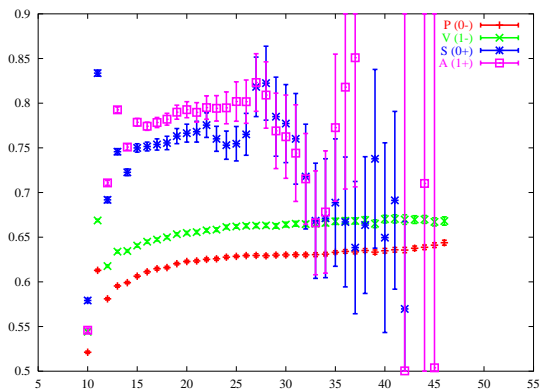}
\end{center}
\caption{Effective mass of the low-lying heavy light mesons with spin-parity \(J^{P}=0^{\mp}\) and \(1^{\mp}\), with \(m_{\rm heavy}a=0.3\) and \(m_{\rm light}a=0.040\).}
\label{fig:effectivemass}
\end{figure}

In Figure \ref{fig:effectivemass} we present typical effective mass for the four \(J^{P}\) states we are discussing.
\noindent
Similarly reasonable plateaux are obtained for heavy quark masses of 0.1, 0.2, 0.3 and 0.4, and light quark masses of 0.016, 0.024, 0.032 and 0.040.  They allow meson mass extraction for all the spin-parity combinations of \(J^{P}=0^{\mp}\) and \(1^{\mp}\).  The pseudoscalar (\(0^{-}\)) and vector (\(1^{-}\)) mass values are obtained with the light quark mass of 0.008 as well.

Using the known strange mass of \(m_{\rm strange}a=0.0295(14)\) \cite{Aoki:2005ga}, the calculated pseudoscalar heavy light masses with \(m_{\rm heavy}a=0.3\) and 0.4, and the experimental \(D_{s}(0^{-})\) mass of 1968.3(5) MeV, we set the charm mass as \(m_{\rm charm}a=0.360(7)\) by interpolation. With these values we interpolate to obtain \(D_{s}\) states masses and extrapolate to \(D\) states ones.  The results are summarized in Table \ref{tab:calculatedmass},
\begin{table}
\begin{center}
\hfill
\begin{tabular}{lll}
\multicolumn{1}{c}{\(J^{P}\)}&\multicolumn{1}{c}{this calculation (MeV)}\\
\hline\hline
\(D^{\pm}(0^{-})\)&1876.7(1.2) (0.4\% over)\\
\(D^{*\pm}(1^{-})\)&1968(2) (2\% under)\\
\(D^{*}_{0}(0^{+})\)&2362(20) (2.5\% over)\\
\(D'_{1}(1^{+})\)&2455(17) (1.2\% over)\\
\end{tabular}
\hfill
\begin{tabular}{ll}
\multicolumn{1}{c}{\(J^{P}\)}&\multicolumn{1}{c}{this calculation (MeV)}\\
\hline\hline
\(D^{\pm}_{s}(0^{-})\)&1968.3 (input)\\
\(D^{*\pm}_{s}(1^{-})\)&2055(6) (2.7\% under)\\
\(D^{*\pm}_{s0}(0^{+})\)&2380(30) (2.5\% over)\\
\(D^{*\pm}_{s1}(1^{+})\)&2460(40) (0.03\% over)\\
\end{tabular}
\hfill
\end{center}
\caption{Calculated mass of \(D_{s}\) (right, by interpolations for both charm and strange) and \(D\) (left, by interpolation for charm and extrapolation for non-strange) states with \(m_{\rm charm}=0.360\) and \(m_{\rm strange}=0.0295\) in lattice units.  ``over'' or ``under'' refers to the difference from the experimental values.  Statistical errors only.}
\label{tab:calculatedmass}
\end{table}
and are in reasonable agreement with the experiments, to within a few \%.

Parity splittings for the strange \(D_{s}\) states turn out as \(\Delta_{s, 0} = 410(30)\) MeV, about 18\% over estimate compared with 349 MeV experiment, and \(\Delta_{s, 1} = 380(20)\) MeV, about 10\% over estimate compared with 347 MeV experiment.  These calculated values are degenerate within statistical errors.  They are also closer to the experiments than previous lattice results.
For the non-strange \(D\) states they are \(\Delta_{s, 0} = 487(20)\) MeV, about 11\% over estimate compared with 439 MeV experiment, and \(\Delta_{s, 1} = 490(17)\) MeV, about 18\% over estimate compared with 417 MeV experiment.  Again they are statistically degenerate and are closer to the experiments than the previous lattice results.

In contrast the hyperfine splittings are not so well reproduced:  about 87(6) MeV for \(D_{s}\) states, reproducing about 60\% of the 144 MeV experiment, and about 91(2) MeV for \(D\) states, reproducing about 65\% of the 141 MeV experiment.

\begin{figure}
\begin{center}
\begin{minipage}[t]{0.45\textwidth}
\begin{center}
\includegraphics[width=.9\textwidth]{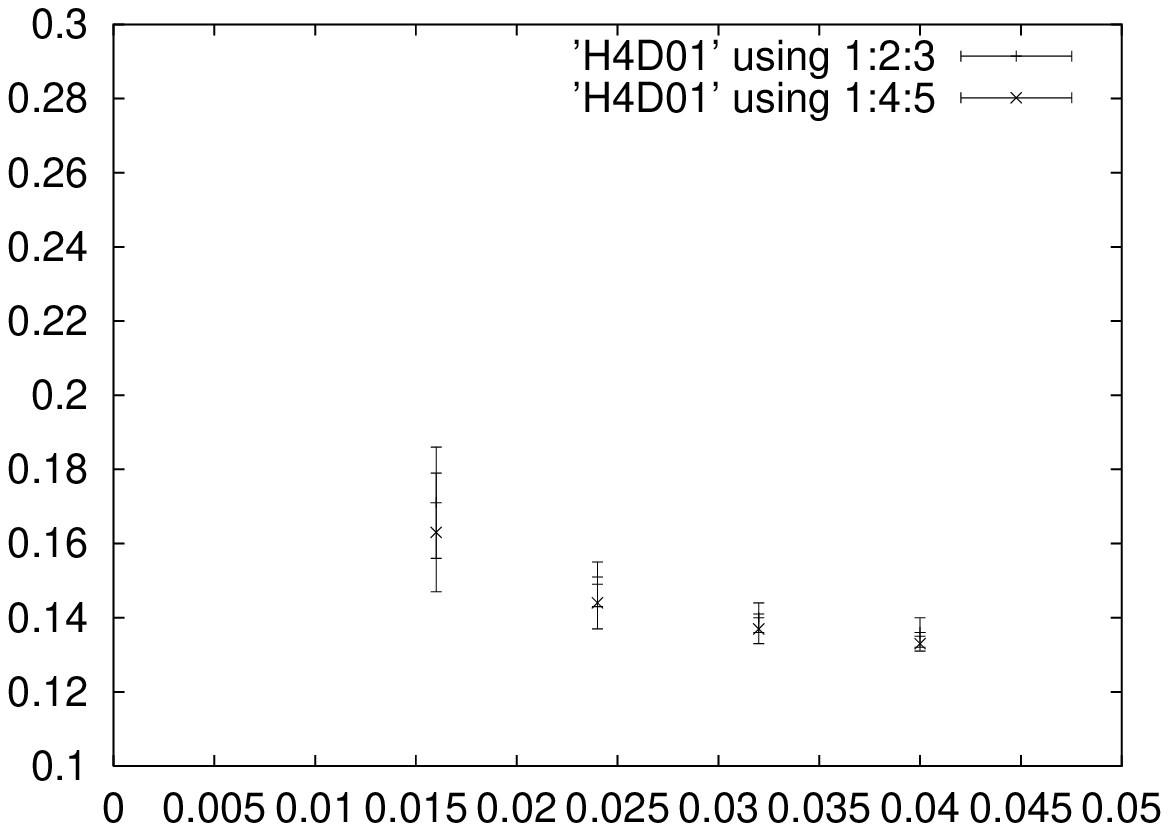}
\end{center}
\end{minipage}
\begin{minipage}[t]{0.45\textwidth}
\begin{center}
\includegraphics[width=.9\textwidth]{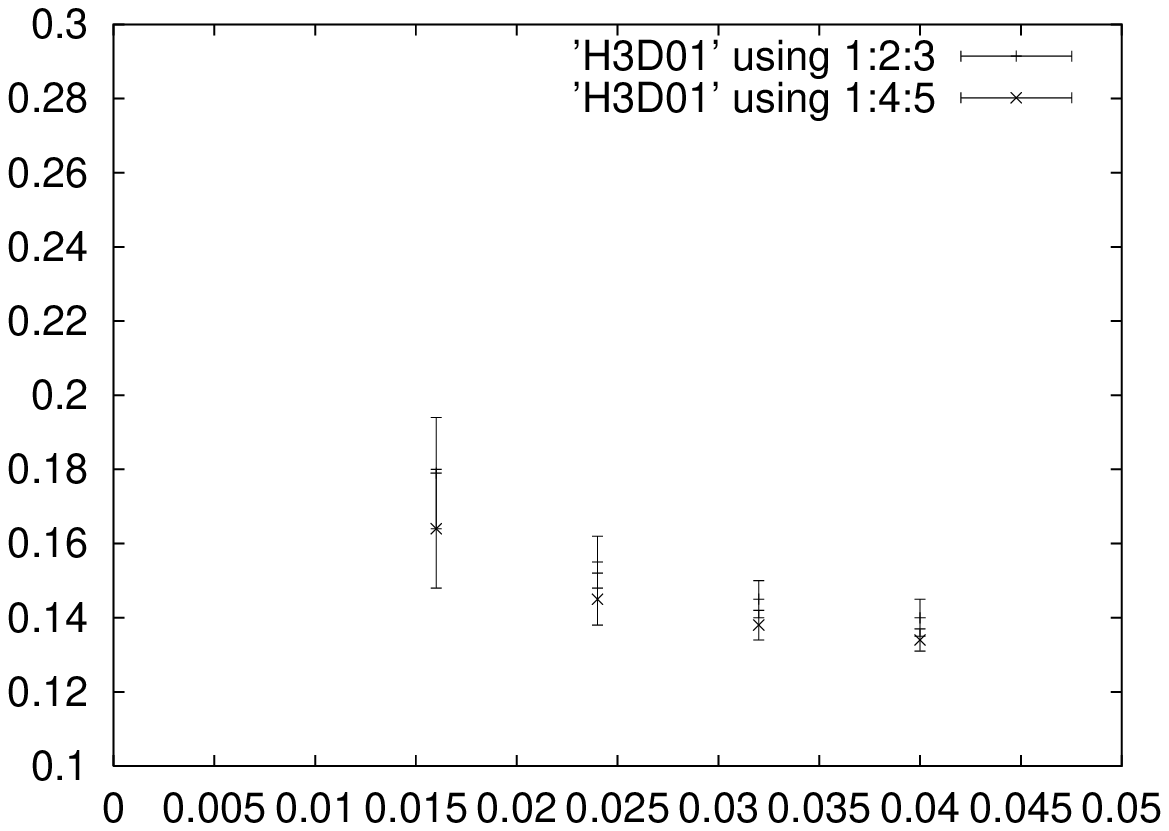}
\end{center}
\end{minipage}
\begin{minipage}[t]{0.45\textwidth}
\begin{center}
\includegraphics[width=.9\textwidth]{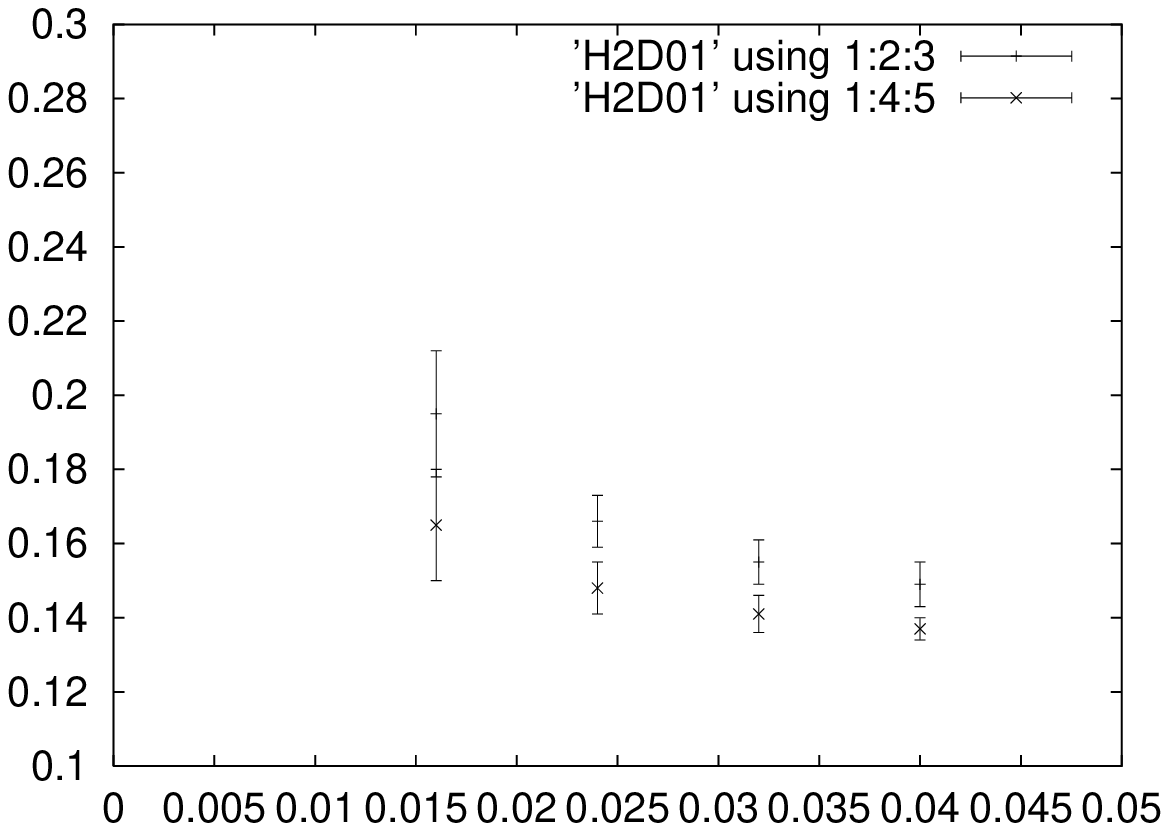}
\end{center}
\end{minipage}
\begin{minipage}[t]{0.45\textwidth}
\begin{center}
\includegraphics[width=.9\textwidth]{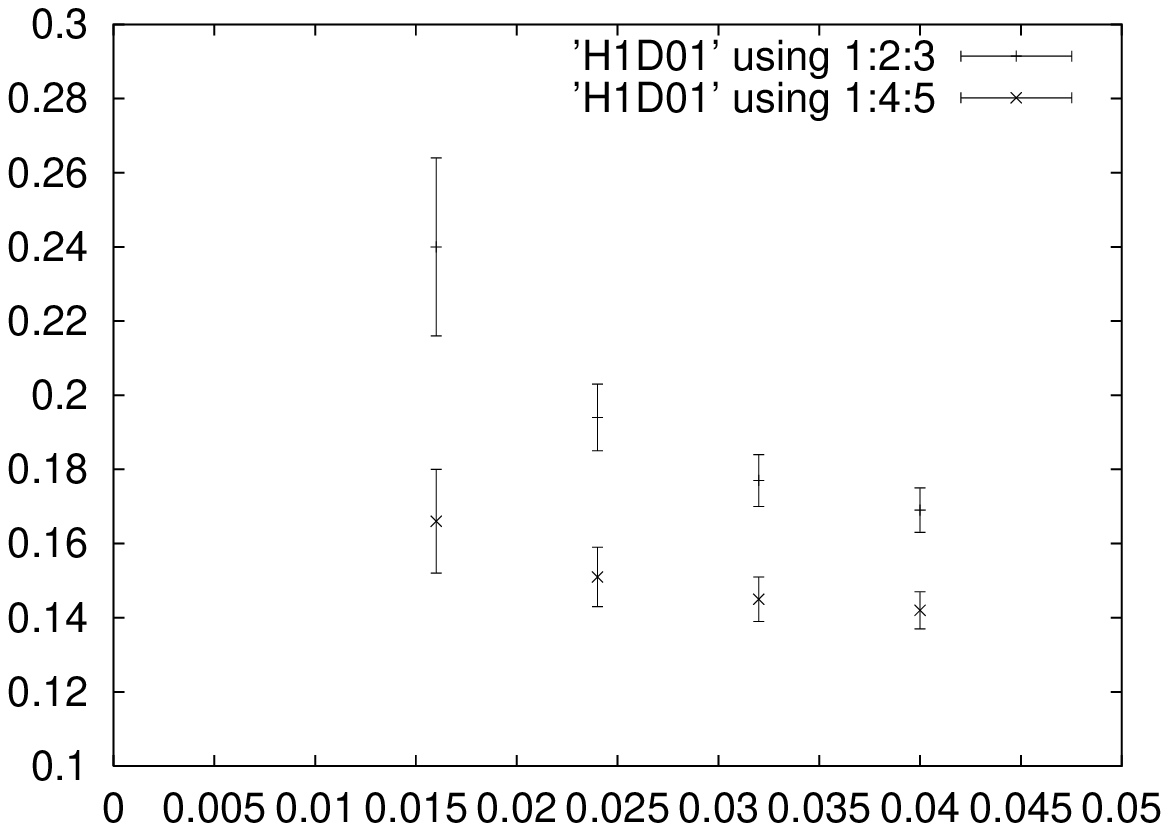}
\end{center}
\end{minipage}
\end{center}
\caption{Quark mass dependences of the parity splittings, \(\Delta_{J=0}\) (\(+\)) and \(\Delta_{J=1}\) (\(\times\)), with \(m_{\rm heavy}a=0.4\) (upper left panel), 0.3 (upper right), 0.2 (lower left), and 0.1 (lower right).  Each panel shows dependence on the light quark mass.}
\label{fig:massdependence}
\end{figure}

The quark mass dependences of parity splittings are summarized in Figure \ref{fig:massdependence}.
\noindent
They are statistically indistinguishable for \(m_{\rm heavy}a=0.4\) and 0.3, but \(\Delta_{J=0} > \Delta_{J=1}\) for lighter \(m_{\rm heavy}\) as \(\Delta_{J=0}\) increases while \(\Delta_{J=1}\) stays more or less constant.  The latter behavior may be supported by \(K_{1}(1270) - K^{*}(892)\).  Both \(\Delta_{J=0}\) and \(\Delta_{J=1}\) increase toward lighter \(m_{\rm light}\).

\section{Conclusions}

We are studying charm-light meson systems with all the quark flavors described by DWF and with quenched DBW2 gauge ensemble at lattice cutoff of about \(a^{-1}\sim\mbox{\rm 3 GeV}\).  Our preliminary conclusions are:   DWF well describe charm on the quenched DBW2 ensemble at this cutoff.   With interpolations to bare quark mass values of \(m_{\rm strange}a=0.0295\) and \(m_{\rm charm}a=0.360\) in lattice units, the masses of the  \(J^{P}=0^{\pm}\) and \(1^{\pm}\) \(D\), \(D^{*}\), \(D_{s}\) and \(D_{sJ}\) states are well reproduced to within a few \%; their parity splitting, \(\Delta_{J}\), are better reproduced than previous works, with only 10-20 \% over estimations; the experimental observation of \(\Delta_{ud}>\Delta_{s}\) is reproduced; but the hyperfine splittings are only 60-65 \% reproduced.  Regarding the dependence on heavy quark mass, \(\Delta_{J=0}\) and \(\Delta_{J=1}\) are degenerate for \(m_{\rm heavy}a >\) 0.2-0.3; \(\Delta_{J=0}\) increases as \(m_{\rm heavy}\) decreases further while \(\Delta_{J=1}\) does not.

We thank RIKEN, Brookhaven National Laboratory and the U.S. Department of Energy for providing the facilities essential for the completion of this work.

\end{document}